# Supersymmetric microring laser arrays

Bikashkali Midya[1], Han Zhao[2], Xingdu Qiao[1], Pei Miao[1], Wiktor Walasik[3], Zhifeng Zhang[2], Natalia M. Litchinitser[3], Liang Feng[1]

[1]Department of Materials Science and Engineering, University of Pennsylvania, Philadelphia, PA 19104, USA
[2]Department of Electrical and Systems Engineering, University of Pennsylvania, Philadelphia, PA 19104, USA
[3]Department of Electrical and Computer Engineering, Duke University, Durham, North Carolina 27708, USA

Coherent combination of emission power from an array of coupled semiconductor lasers operating on the same chip is of fundamental and technological importance. In general, the nonlinear competition among the array supermodes can entail incoherence and spectral broadening, leading to spatiotemporally unstable and multimode emission pattern and thus poor beam quality. Here, by harnessing notions from supersymmetric (SUSY) quantum mechanics, we report that the strategic coupling between a class III-V semiconductor microring laser array with its dissipative *superpartner* can be used to limit the number of supermodes available for laser actions to one. We introduce a novel approach based on second-order SUSY transformation in order to drastically simplify the superpartner array engineering. Compared to a conventional laser array, which has multimode spectrum, a SUSY laser array is observed to be capable of operating in a single (transverse) supermode. Enhancement of the peak output intensity of the SUSY laser array has been demonstrated with high-efficiency and lower lasing threshold, compared with a single laser and a conventional laser array. Our experimental findings pave the way towards broad-area and high-power light generation in a scalable and stable fashion.

## Introduction

Supersymmetric (SUSY) quantum field theory predicts that each elementary particle and antiparticle of the standard model has its own partner—a superpartner (or sparticle)—endowed with opposite statistics [1,2]. A space-time symmetry transformation between a particle and its superpartner theoretically resolves the puzzle of the Higgs mass renormalization (the hierarchy problem) in particle physics beyond the standard model; the loop contributions of one particle to the Higgs field are exactly cancelled by the contributions of its superpartner, thus confirming the observed mass [3]. While the search for sparticles is being continued, the ramification of the SUSY theory has attracted attention in other branches of physics [4]. In optics, for example, the isomorphism between Schrödinger equation and the scalar Helmholtz equation facilitates to utilize the mathematics of SUSY for the strategic manipulation of light scattering, guiding, and localization in novel photonic materials and structures. In the framework of unbroken SUSY, starting from a given primary optical structure it is possible to construct its superpartner with identical eigen-spectrum apart from the fundamental mode which is deleted by the SUSY transformation [5-7]. When the former structure is judiciously coupled with its unbroken superpartner, the global mode matching of the combined structure leverages efficient mode conversion and multiplexing in passive configurations [8], and strategic selection of a single transverse supermode in an active laser array [9,10].

Integrated semiconductor laser arrays that consist of a large number of densely packed emitters is of special importance for next generation of applications such as material processing, broad-area displays, industrial heating, and lidars [11-13]. The coherent combination and phase locking are essential in obtaining higher power output from a large transverse area than those available from a single semiconductor laser. This has offered advantages of narrowing the spatial extent of the radiation beam while maintaining a narrower output spectrum [14-18]. However, a laser array, in general, operates on many closely spaced transverse modes (supermodes) of the same longitudinal order; the degeneracy of each cavity longitudinal mode is lifted owing to the proximity coupling among the individual cavities [Fig.1(a)]. Nonlinear modal competition due to limited gain often results in pulsation and filamentation that degrade the spectral, spatial, and temporal characteristics of the radiation [19,20]. While the single longitudinal mode lasing was addressed by parity-time symmetry [21], novel strategy and symmetry consideration are required to support single transverse supermode lasing. Sophisticated laser-array designs based on gain tailoring, special diffractive or non-Hermitian coupling are proposed to mitigate the complexity of transverse supermode lasing [22-24].

In this article, we theoretically analyze and experimentally demonstrate single (transverse) supermode lasing in a SUSY designed array (schematically shown in Fig.1b) of III-V semiconductor active microring resonators. With this aim, we construct a superpartner array whose linear spectrum is identical to that of the main array apart from the fundamental supermode. The operating principle of single-supermode SUSY laser array then follows from the judicious coupling of the main array with its dissipative superpartner, which ensures the reduction of supermode competition by decreasing the number of available lasing modes [9]. The global mode matching between the SUSY partners ensures the suppression of all, but the fundamental, supermodes whose modal loss and thus lasing threshold are increased, owing to the added stronger loss in the superpartner channels [Fig.1(a)]. The fundamental supermode, confined only in the main array lacking a superpartner, has the lowest threshold and is favorable for lasing upon a suitably applied pumping. The advantage of the fundamental-supermode selection is that the emission from all the rings are in phase. Although a laser is a highly nonlinear system, the SUSY linear-mode suppression mechanism mentioned above is observed to persist in our experiment. Compared to a conventional laser array, which is highly multimoded, our SUSY array is demonstrated to support a single supermode lasing corresponding to two nearby longitudinal orders. Apart from its single-mode property, the SUSY laser array is observed to be highly efficient, requiring lower lasing threshold, and has enhanced intensity compared to a single laser and a conventional laser array at the identical pumping condition.

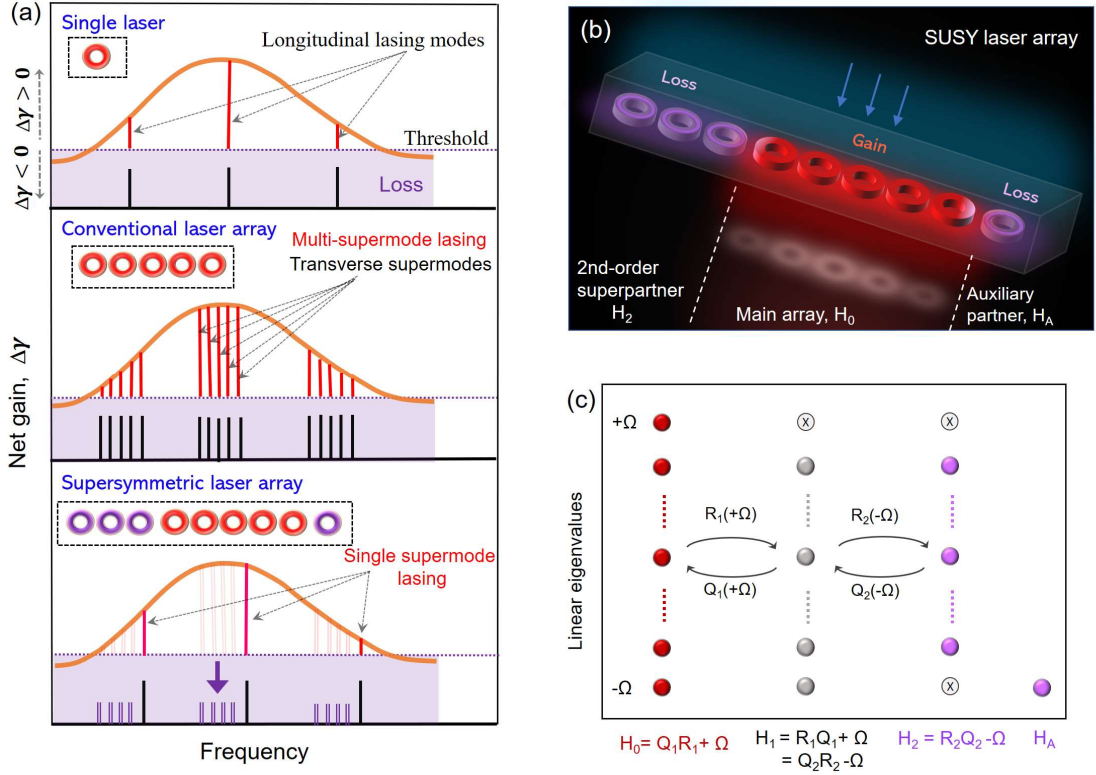

**Fig.1. Fundamental concept behind the single (transverse) supermode lasing.** (a) A single laser supports multiple longitudinal cavity modes (vertical, black lines) separated by free spectral range. When $N$ such rings are coupled, $N$ closely spaced transverse supermodes emerges with equal threshold (middle panel). In the presence of gain, a suitable pump can induce simultaneous lasing of all the supermodes. The global mode coupling with a dissipative superpartner in a SUSY laser array, schematically shown in (b), can increase the threshold of undesired modes (shown in vertical, purple lines in the bottom panel), enforcing a lasing of a single supermode which lacks a superpartner counterpart. (c) The second-order SUSY transformation for the superpartner design.

## Theory

While a superpartner resulting from a single SUSY transformation (deletion of a single eigenmode) was proposed for a single mode SUSY laser array in theory [9], the resulting superpartner elements have small but non-negligible frequency detuning which can be detrimental especially if the coupling between the lasers are much weaker compared to the onsite energies (as is the case of microring lasers) [10]. Here the challenge is overcome by the introduction of a novel design technique based on two consecutive SUSY transformations applied to a homogeneous array enabling elimination of two supermodes with the largest and smallest frequencies, respectively [Fig. 1(c)]. *The second-order superpartner array thus obtained consists of identical elements with the resonance frequency identical to that of a main array element.*

We consider a homogeneously coupled one-dimensional array of $N$ identical laser cavities with the resonance frequency $\widetilde{\omega}$ (considered to be the lowest transverse order), nearest-neighbor coupling strength $k$, cavity linear loss $\gamma_0$, and nonlinear gain $g$. In the absence of gain, the linear modes of the system are given by: $H_0 \varphi^{(0)} = \omega \varphi^{(0)}$, where $H_0$ is a tridiagonal matrix with diagonal entries $\tilde{\omega} - i\gamma_0$, off-diagonal entries $k$, and $\omega_n = \tilde{\omega} - i\gamma_0 + 2k \cos\frac{n\pi}{N+1}$, $n = 1, 2 \ldots N$, are the eigenfrequencies. Note that the normalized eigenvalues $\Omega_n = (\omega_n - \tilde{\omega} + i\gamma_0)/k$ appear to be symmetric with respect to zero, such that $\Omega_n = -\Omega_{N-n+1}$. The first-order superpartner $H_1$ of $H_0$ is obtained by the discrete SUSY transformation (QR-factorization algorithm [10]): $H_0 - \Omega_1 = Q_1 R_1$, $H_1 = R_1 Q_1 + \Omega_1$, which implies that the fundamental frequency is deleted from the otherwise isospectral spectrum of $H_1$ i.e. Spec($H_1$)= Spec($H_0$)\{$\Omega_1$}. A repeated application of the similar procedure at the frequency $\Omega_N = -\Omega_1$, yields a second-order superpartner $H_2$, such that $H_1 - \Omega_N = Q_2 R_2$, and $H_2 = R_2 Q_2 + \Omega_N$. In this case, Spec($H_2$)=Spec($H_1$)\{$\Omega_N$}=Spec($H_0$)\{$\Omega_1, \Omega_N$}, and the linear eigenmodes of $H_0$ and $H_2$ are related by: $\varphi^{(0)}(\Omega_n) = Q_1 Q_2 \varphi^{(2)}(\Omega_n)$ and $\varphi^{(2)}(\Omega_n) = R_2 R_1 \varphi^{(0)}(\Omega_n)$. This implies that a first-order superpartner is spectrally equivalent to a pair of subarrays, ($H_2$, $H_A$), containing the second-order superpartner and an isolated auxiliary partner $H_A$ (which consists of a single element of frequency $\Omega_N$). The coupling of the main array with the lossy subarrays $H_2$ and $H_A$ [as shown in Fig.1(b) and 2(a)] then ensures energy dissipation of all the isospectral supermodes, while the fundamental supermode remains unaffected provided that the total coupling between the arrays is much weaker than $k$. The SUSY laser array, so constructed, thus consists of $2N-1$ lasers—$N$ belongs to the main array, $N-2$ to $H_2$ and a single laser belongs to the $H_A$. The most remarkable property of $H_2$ is that all of its elements have *zero relative frequency detuning*, while the couplings between the elements remain inhomogeneous. The second-order SUSY technique thus drastically simplifies the large-scale laser array engineering, as all the laser elements, but the auxiliary one, are identical in geometry.

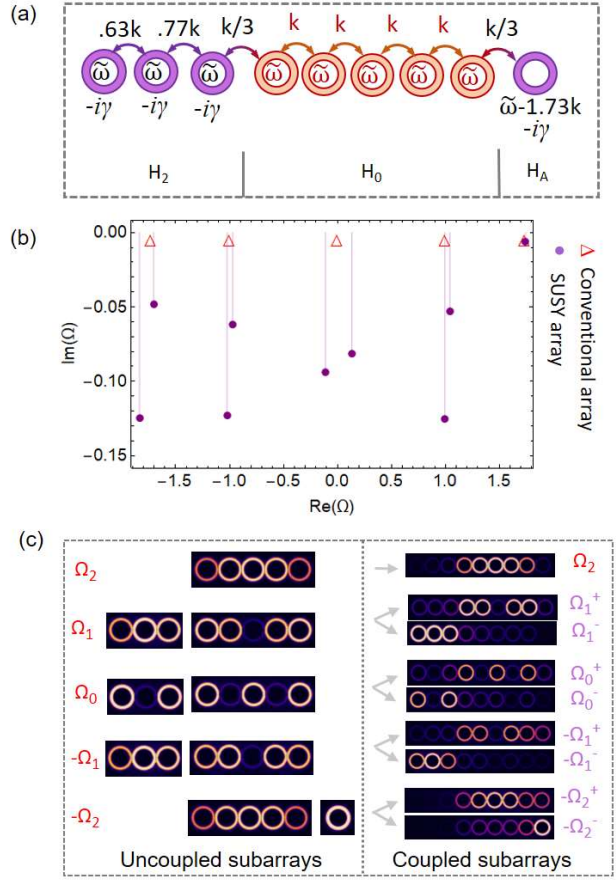

**Fig.2. SUSY laser array linear mode analysis.** (a) Design parameter of a SUSY microring laser array containing nine coupled lasers. (b) and (c) showing the linear eigen spectrum and corresponding modal intensities (only the cross-sectional view is shown), respectively. The spectrum shows that the fundamental mode of the SUSY array has least threshold. All other modes of the conventional laser array splits into symmetric and anti-symmetric pairs $\Omega^{\pm}$. Here $\tilde{\omega} = 0$, $\gamma_0 = 0.005\,k$, $\gamma = k/6$ and $k=1$ are considered.

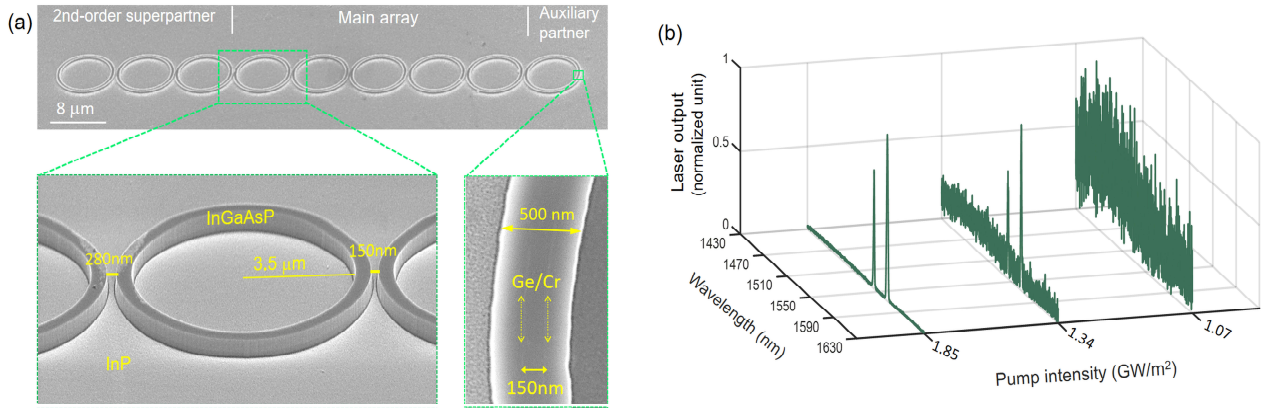

**Fig.3. Device and spectrum.** (a) Scanning electron microscope images of the fabricated SUSY microring laser array. The image was taken before the transfer of sample into a silica substrate. (b) Evolution of the photon emission spectrum from photoluminescence to amplified spontaneous emission and then to supermode lasing at the wavelengths of about 1544 nm, and 1568 nm, when the peak intensity of pump light is increased. The two lasing peaks, separated by 24 nm, belong to two longitudinal modes of different azimuthal order in a single microring laser [see Fig. 4a].

To experimentally validate and compare the spectral responses of the above-mentioned SUSY laser array, we have designed a single microring laser, a conventional array of five lasers (without superpartner), and the SUSY array of nine lasers. The design parameters and the corresponding linear eigenmode analysis are shown in Fig.2. The cavity intrinsic loss $\gamma_0$ is considered equal for all the rings (in main as well as superpartner array), while additional loss $\gamma$ is introduced in the superpartner lasers. To avoid the mode decoupling between the main and superpartner subarrays, the loss contrast must be less than the coupling between the subarrays. As seen from the Fig.2(b), the conventional laser array has five eigenfrequencies each with equal amount of loss, whereas the SUSY laser array possesses enhanced losses to all the modes except for the largest frequency. Finite element simulations were performed to obtain the corresponding Transverse-electric polarized field intensity distributions in the InGaAsP laser array in a silica substrate. The additional loss is estimated by considering a thin layer of Cr/Ge on the ring structures, and the coupling between the lasers are tailored by adjusting the center to center distance between the rings. The results are shown in Fig.2(c). Apart from the fundamental supermode that is predominantly confined in the main array, other supermodes have significant overlap with the lossy superpartner subarrays, leading to the single transverse mode lasing.

## Experiment

The devices were fabricated by electron-beam lithography and plasma etching on 200 nm thick InGaAsP multiple quantum wells as the gain material with the gain bandwidth spanning from 1430 nm to 1600 nm. The scanning electron microscope images of the SUSY array sample with the crucial geometric parameters are shown in Fig.3(a). To ensure the loss required in the superpartner array, and to properly match the frequency detuning in the auxiliary ring, thin layers of Cr/Ge were appropriately deposited on the corresponding rings through aligned overlay electron beam lithography and lift-off. After using plasma-enhanced chemical vapor deposition to deposit a 3μm thick layer of $SiO_2$ for protecting the sample. The InGaAsP ring lasers on the InP wafer were then bonded with glass slide, followed by a selective wet etching using $HCl/H_3PO_4$ to remove only the InP substrate. Each ring is 4 μm in outer-radius, 500 nm in width and 200 nm in height. Coupling between the rings in the main array, placed 150 nm distance apart, was experimentally measured to be 105 GHz around 1550 nm wavelength. The gaps between the rings in $H_2$ were 200 and 225 nm, while the gaps between the main and partner subarrays were 280 nm.

The single rings and the ring arrays were pumped optically by a nanosecond pulsed laser with a 50 kHz repetition rate and an 8 ns pulse duration at the wavelength of 1064 nm. To uniformly pump each elementary ring in the ring array, the transverse profile of the pumping laser beam was shaped in a linear manner by a cylindrical lens and a near-infrared (NIR) objective with 0.4 numerical aperture. The emissions of light oscillating in the laser array were collected by another NIR long working-distance objective whose spectra were subsequently measured by a monochromator. The emission spectrum of a SUSY array is shown in Fig.3(b). Two lasing peaks around 1544 nm and 1568 nm wavelengths emerge when pump densities are gradually increased. The comparison of the emission spectrum with that of a single element laser and a conventional five ring array under the same pump intensity is shown in Fig.4 (a). On the one hand, the spectrum of SUSY arrangement appears almost indistinguishable from the single ring narrow spectrum, implying that the SUSY ring supports single supermode lasing. On the other hand, the spectrum of a conventional five ring array, as expected, shows multiple peaks, which reflects the mode competition between the supermodes. The imaged lasing emission profiles from each of these cases are shown in the inset of Fig.4(a). From the image of the SUSY array, we observe the light radiation only from the five main rings, the presence of the superpartner rings only serve to suppress unwanted mode. Although the conventional laser array possesses a total emission power comparable to the SUSY array, the peak intensity produced by this SUSY laser is significantly enhanced owing to the single supermode nature. The characteristic light-light curve for a single element laser and the SUSY laser array are shown in Fig.4(b). The lasing intensity produced by this SUSY laser is over 4 times higher than that from the single elementary laser at the 2.46 GW/m$^2$ pump density. The much steeper slope implies that the SUSY array is highly efficient compared to a single ring, indicating that the large transverse gain in the former case increases overall efficiency.

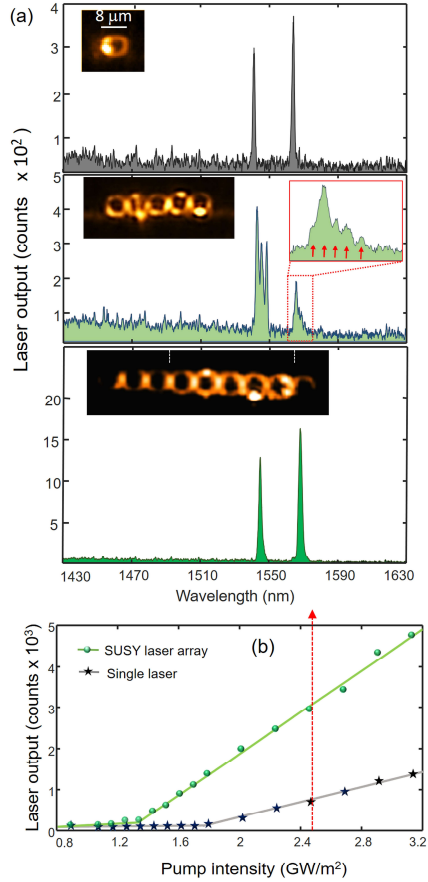

Fig.4. **Comparison of measured spectra**. (a) Output spectrum for a single laser (top panel), a conventional laser array (middle panel) and the SUSY laser array (bottom panel) when 2.46 GW/m$^2$ pump intensity is considered. The conventional array is seen to be highly multimoded (as evident from the magnified spectrum shown in inset). The spectrum of single laser and the SUSY laser array are almost indistinguishable. The enhancement of lasing output peak intensity is seen in the SUSY laser array. The nearfield images of lasing emissions are also shown. (b) Light-light curve showing the lowering of the threshold and enhancement of lasing output in a SUSY laser array compared to a single microring laser.

## Conclusion

We report the first experimental observation of a highly-efficient, low-threshold, broad-area light generation from a microring laser array designed according to the principle of unbroken SUSY. The laser array is demonstrated to operate predominantly in a single transverse supermode of two adjacent longitudinal orders. The elegant strategy of dissipative coupling between the SUSY partner arrays is validated to suppress undesired modes in favor of fundamental mode lasing where all individual laser cavities oscillate in phase. The notion of second-order superpartner introduced here is conducive for large-scale SUSY laser array design. Our experimental results open the door for high-power light source. It would be worthwhile to explore the interplay between parity-time- and super-symmetry in a large-scale laser array to tailor the high-power light emission restricted to a single longitudinal as well as single transverse mode for ultimate spectral purity.

*Note added*. The preprint [25] reported a similar experiment on an active waveguide array.

*Acknowledgement*. This work was carried out in part at the Singh Center for Nanotechnology, part of the National Nanotechnology Coordinated Infrastructure Program, which is supported by the National Science Foundation grant NNCI-1542153.

*Funding*. Army Research Office (W911NF-17-1-0400), National Science Foundation (CMMI-1635026; IIP-1718177; CBET-1706050).

## Sample preparation

The microring resonator array is fabricated using electron-beam lithography and plasma etching techniques. Hydrogen silsesquioxane (HSQ) solution in methyl isobutyl ketone (MIBK) is used as negative electron beam lithography resist. The concentration ratio of HSQ (FOX15) and MIBK is adjusted such that after exposure and developing a layer of resist that is thick enough as an etching mask and an 150 nm gap between microring resonators can be achieved. The resist is then soft baked and ring arrays are written by electron beam exposure. Electrons convert HSQ resist to an amorphous structure similar to $SiO_2$. The patterned wafer is then immersed and slightly stirred in tetramethylammonium hydroxide (TMAH) solution (MFCD-26) for 120 seconds and rinsed in DI water for 60 seconds. The exposed and developed HSQ serves as a mask for the subsequent inductively coupled plasma etching process that uses $BCl_3$ : Ar plasma with gas ratio of 15 : 5 sccm respectively and runs with RF power of 50W and ICP power of 300W under a chamber pressure of 5mT. After dry etching, HSQ resist is removed by immersing the sample in buffered oxide etchant (BOE) for 60 seconds.

Polymethyl methacrylate (PMMA A2) is used as positive resist for the subsequent registration electron beam exposure [Fig.5(b)]. The patterned wafer is immersed and stirred in MIBK: IPA developer and rinsed in IPA. Metal is then deposited with electron-beam physical vapor deposition tool. After deposition, wafer is immersed in PG remover at 60°C for lift-off process. The registration exposure and the following steps are carried out once more for the deposition of metal with another designed thickness for the auxiliary laser elements. In the next step [Fig.5(c)], a $SiO_2$ (silicon dioxide) coating is applied by means of PECVD. The wafer is then bonded to a glass slide which functions as a holder. Finally, the InP substrate is removed by wet etching with a mixture of HCl (Hydrochloride acid) and $H_3PO_4$ (Phosphoric acid).

Figure 6 shows the experimental setup for the characterization and emission spectrum measurement of the laser arrays.

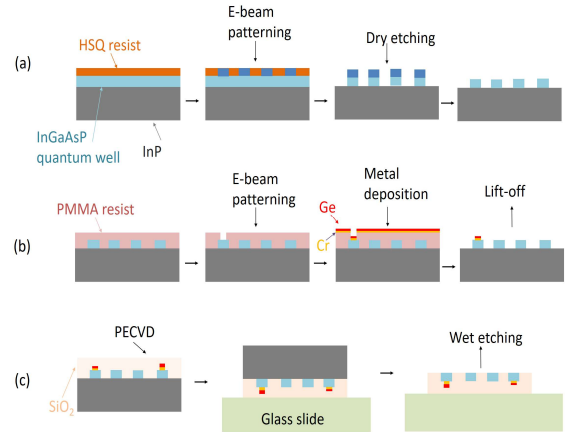

fig. 5. Flowchart showing the fabrication processes of SUSY microring laser array.

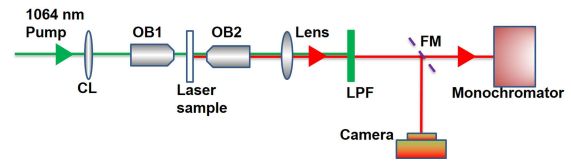

fig. 6. Experimental setup for the laser characterization. The laser sample is optically pumped by a pulsed laser at the wavelength of 1064 nm. The linearly uniform pumping pattern is shaped by a cylindrical lens and then focused on the sample with a near-infrared long working distance objective. By manipulating the flip mirror (FM), the collected laser emission is imaged to the camera, or enters the monochromator for spectral measurement. CL: cylindrical lens; OB1, OB2: near-infrared long-working distance objectives. LPF: long pass filter with 1450 nm cut-off wavelength.

## Stability analysis

We numerically verify the nonlinear emission dynamics of the SUSY laser array. Assuming very fast carrier dynamics, the dynamical equation governing the intensity evolution of the lasers can be considered as

$$i\frac{d\varepsilon_j}{dt} = \left(\omega_j - i\gamma_j + \frac{i\,g}{1+\alpha|\varepsilon_j|^2}\right)\varepsilon_j + k_{j-1}\varepsilon_{j-1} + k_j\varepsilon_{j+1}$$

$j$=1,2,3 and 9 corresponds to superpartner lasers, and $j$=4,…8 corresponds to those in the main array [see Fig.2(a) of the main text]. Here, $\alpha$ is the gain saturation term and considered to be 1 in the simulation, and the gain $g$= 0.18$k$. The loss term $\gamma_j$ is equal to $\gamma_0$ for the main array elements, and is $\gamma + \gamma_0$ for the superpartner lasers. Other parameters remain same as in Fig. 2(a) of main text. As shown in figure 7(b), the laser intensities in the SUSY array are temporally stable after an initial transient period is elapsed. Whereas, the conventional laser array (in the absence of superpartner) reveals highly chaotic dynamics as shown in Fig. 7(a). The phase locking between the SUSY array lasers is shown in Fig.7(c).

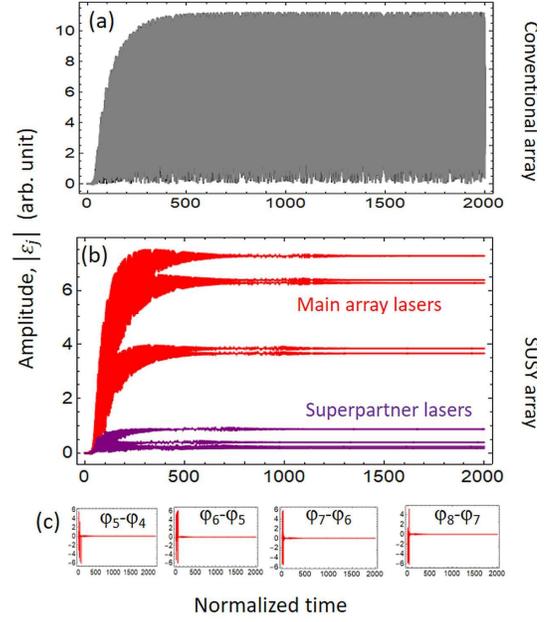

Fig.7. Intensity evolution of the lasers in a conventional laser array (a), and in a SUSY laser array (b). (c) The relative phase between the main array elements of the SUSY array.